\documentclass[a4paper,11pt]{article}
\usepackage{jheppub}
\usepackage{slashed}
\usepackage{color}
\usepackage{rotating}
\usepackage{cancel}

\usepackage{graphicx}
\usepackage{dcolumn}
\usepackage{bm}

\newcommand{\CA}{{\rm C}_{\rm A}}
\newcommand{\CF}{{\rm C}_{\rm F}}

\newcommand{\gcusp}{\gamma_{\rm cusp}}

\newcommand{\taun}{\mathcal{T}_N}
\newcommand{\tauzero}{\mathcal{T}_{0}}
\newcommand{\taucut}{\tau^{\rm cut}}

\title{\boldmath Next-to-Next-to-Leading Order  $N$-Jettiness Soft Function for One Massive Colored Particle Production at Hadron Colliders}

\author[a]{Hai Tao Li}
\author[b,c]{and Jian Wang}

\affiliation[a]{ARC Centre of Excellence for Particle Physics at the Terascale, School of Physics and Astronomy, Monash University, VIC-3800 Australia }
\affiliation[b]{PRISMA Cluster of Excellence $\&$ Mainz Institute for Theoretical Physics,
Johannes Gutenberg University, D-55099 Mainz, Germany}
\affiliation[c]{Physik Department T31,  Technische Universit\"at M\"unchen, James-Franck-Stra\ss e~1,
D--85748 Garching, Germany}

\emailAdd{haitao.li@monash.edu}
\emailAdd{j.wang@tum.de}

\abstract{The $N$-jettiness subtraction has proven to be an efficient method to perform differential QCD next-to-next-to-leading order (NNLO) calculations
in the last few years. One important ingredient of this method is the NNLO soft function.
We calculate this soft function for one massive colored particle production at hadron colliders.
We select the color octet and color triplet cases to present the final results.
We also discuss its application in NLO and NNLO differential calculations.}

\arxivnumber{1611.02749}

\preprint{CoEPP-MN-16-28 \\   \rightline{MITP/16-112} \\   \rightline{TUM-HEP-1067/16} }

\begin{document}
\maketitle
\flushbottom

\section{Introduction}
After the discovery of the Higgs boson, one of the main tasks of the Large Hadron Collider~(LHC) is to discover new physics beyond the Standard Model (SM).
So far there has been no sign of new physics at the LHC. In the future,
it is possible that the new physics would appear in small deviations from the SM predictions.
Therefore, along with the accumulating experimental data,
accurate and precise theoretical predictions for the SM processes are mandatory for discovering new physics.
Including higher-order effects in quantum chromodynamics~(QCD) perturbation theory can help to improve the  theoretical predictions.
More specifically, one needs to consider fixed-order calculations for general observables, and analytical resummation or parton shower simulations to resum large logarithmic terms for certain delicately designed observables.

Soft-collinear effective theory (SCET) ~\cite{Bauer:2000ew,Bauer:2000yr,Bauer:2001ct,Bauer:2001yt,Beneke:2002ph}
has been widely used in analytical resummation and fixed-order calculations; see Ref.~\cite{Becher:2014oda} for a review. One of recent progresses is the application of SCET to studying the $N$-jettiness event shape variable $\mathcal{T}_N$~\cite{Stewart:2010tn}.
When $\mathcal{T}_N \to 0$, the cross section is factorized as the convolution of a hard function, two beam functions for initial states, a soft function and jet functions \cite{Stewart:2009yx,Stewart:2010tn},
\begin{align}
\label{eq:sigma}
    \frac{d\sigma}{d\taun} \propto \int H\otimes B_1 \otimes B_2 \otimes S \otimes \bigg(\prod_{n=1}^{N}J_n \bigg)~,
\end{align}
where the $N$-jettiness event shape variable is defined as~\cite{Stewart:2010tn}
\begin{align}
\label{eq:tau}
\taun = \sum_k \min_i\left\{n_i \cdot q_k \right\}~.
\end{align}
Here $n_i$ $(i=a,b,1,...,N)$ are light-like reference vectors representing the moving directions of massless external particles,
and $q_k$ denotes the momentum of soft or collinear radiations.
The hard function $H$ encodes all the short distance information and can be obtained from the Wilson coefficient when matching full QCD onto SCET. The beam functions $B_i, ~(i=1,2),$   incorporate the effects of  parton distribution functions
as well as initial-state radiations, which are known up to next-to-next-to-leading order~(NNLO)~\cite{Stewart:2010qs,Berger:2010xi,Gaunt:2014xga,Gaunt:2014cfa}.
The jet function $J_n$ describes the final-state jet with a fixed invariant mass and has been calculated at NNLO~\cite{Becher:2006qw,Becher:2010pd}. The soft function $S$ describes soft interactions between all colored particles.
It has been studied up to NNLO for massless processes recently~\cite{Jouttenus:2011wh,Kelley:2011ng,Monni:2011gb, Boughezal:2015eha,Kang:2015moa}.
Based on these ingredients, the resummed $N$-jettiness distribution can be obtained
\cite{Stewart:2009yx,Stewart:2010pd,Berger:2010xi,Kang:2012zr,Jouttenus:2013hs,Kang:2013nha,Kang:2013lga,Kang:2015swk,Alioli:2015toa}.
Besides the analytical resummation, the $N$-jettiness subtraction method,
based on the expansion of Eq.(\ref{eq:sigma}) in series of the strong coupling $\alpha_s$,
has been proposed to perform NNLO calculations \cite{Boughezal:2015dva,Boughezal:2015aha,Gaunt:2015pea},
and been utilized to provide differential predictions for processes with final-state jets at a hadron collider
\cite{Boughezal:2015ded} and processes at an electron-hadron collider~\cite{Berger:2016inr,Abelof:2016pby}.

In principle, the extension to massive external particles is possible and has been mentioned in Ref.~\cite{Gaunt:2015pea}.
Because there are no collinear singularities along the moving direction of the heavy particle,
the  definition of $\taun$ in Eq.~(\ref{eq:tau}) can control all the infrared (both soft and collinear) singularities.
As a result, there is no need to introduce extra $N$-jettiness axes for heavy particles,
such as in the case of top quark pair production~\cite{Zhu:2012ts,Li:2013mia,Catani:2014qha}
and single top production and decay \cite{Gao:2012ja,Berger:2016oht}.
The effect of including heavy external particle amounts to modifying only the soft function in Eq.~(\ref{eq:sigma}).
We present in this work a calculation of the soft function for one massive colored particle production up to NNLO.

We notice that the soft function for such a process in the energy threshold has already been obtained in Refs.~\cite{Idilbi:2009cc,Beneke:2009rj,Czakon:2013hxa},
which would be useful for threshold resummation.
However, the measurement function for soft function in the $N$-jettiness factorization is different, as indicated by Eq.~(\ref{eq:tau}).
Moreover, in threshold resummation, there are no beam functions, which means the collinear-radiation effects from the initial states are not taken into account in the factorized cross section.
So it can not be directly used in fixed-order calculations.

This paper is organized as follows.
In the next section we briefly introduce the definition of the soft function for one massive colored particle production at hadron colliders.
In section~\ref{sec:Renor}, we derive the structure of the soft function from its properties under renormalization group evolution.
Then, in sections~\ref{sec:nlo} and~\ref{sec:nnlo}, we present the results of next-to-leading order (NLO)
and NNLO soft functions, respectively.
With these results at hand, we discuss its application to NLO and NNLO calculations in section \ref{sec:app}.
We conclude in section~\ref{sec:conc}.
In the appendixes, we show the soft function for a color-triplet fermion production
and results of master integrals in virtual-real corrections.

\section{\label{sec:factor}Definition of the soft function}

In this section we discuss the factorization of the cross section for one massive colored particle production
and present the definition of soft function.

We consider the process
\begin{align}
    P_1 + P_2 \to Q + X~,
\end{align}
where $P_1$ and $P_2$ denote incoming hadrons, $Q$ represents the massive colored particle,
and $X$ includes any inclusive hadronic final state.
The massive colored particle can be massive quarks in the SM or some new particles in extensions of the SM.
For quark-antiquark initial state, the partonic process at leading order (LO) is
\begin{align}
\label{eq:process}
    q(p_1) + \bar{q}(p_2) \to Q(p_3)~.
\end{align}
For later convenience we introduce two light-like vectors
\begin{align}
     n^\mu&=(1,0,0,1), \qquad \bar{n}^\mu=(1,0,0,-1)~.
\end{align}
The momenta given in Eq.~(\ref{eq:process}) can be written as
\begin{align}
     p_1^{\mu} = \frac{m}{2}n^{\mu}~,\quad  p_2^{\mu} = \frac{m}{2} \bar{n}^{\mu}, \quad
     p_3^\mu =\frac{m}{2}( n^\mu+\bar{n}^\mu)~,
\end{align}
where $m$ is the mass of the particle $Q$. The 0-jettiness event shape variable in this process is defined as
\begin{align}
     \mathcal{\tau} \equiv \tauzero = & \sum_{k} \min\{n\cdot q_k, \bar{n}\cdot q_k \}~.
\end{align}

In the limit $\tau \ll m$, the final state contains only one massive particle along with possible soft and collinear radiations.
The cross section in this limit admits a factorized form which can be derived in the frame of SCET.
According to the analysis in Refs.~\cite{Stewart:2009yx,Stewart:2010tn,Stewart:2010qs}, the cross section can be written as
\begin{multline}
\label{eq:fact}
    \frac{d\sigma}{dY d\tau} = \sigma_0 H(\mu^2)\int dt_a dt_b  d\tau_s B_1(t_a, x_a, \mu) B_2(t_b, x_b, \mu)
    \\
    \times S(\tau_s,\mu)\delta\left(\tau-\tau_s - \frac{t_a+t_b}{m}\right)\left(1+\mathcal{O}\left(\frac{\Lambda^2}{m^2},\frac{\tau}{m}\right)\right)~,
\end{multline}
where $\sigma_0$ is the LO partonic cross section, $Y$ is the rapidity of the partonic colliding system in the laboratory frame,
the momentum fractions $x_a=m/\sqrt{s}e^{Y}$ and $x_b=m/\sqrt{s}e^{-Y}$ with $\sqrt{s}$ the colliding energy,
and  $\mu$ is the renormalization scale.
The soft function is defined by the vacuum matrix element
\begin{multline}\label{eq:soft}
    S(\tau, \mu)=\sum_{X_s}
    \Big\langle 0 \Big|\mathbf{\bar{T}} Y_n^{\dagger}Y_{\bar{n}}Y_{v} \Big| X_s \Big\rangle
     \delta\bigg(\tau-\sum_{k} \min\left(n\cdot \hat{P}_k, \bar{n}\cdot \hat{P}_k\right)\bigg)
     \Big\langle X_s\Big| \mathbf{T} Y_nY_{\bar{n}}^{\dagger}Y_{v}^{\dagger}  \Big| 0 \Big\rangle,
\end{multline}
where $\mathbf{T}(\mathbf{\bar{T}})$ is the (anti-)time-ordering operator, and $Y_n$,  $Y_{\bar{n}}$ and $Y_{v}$ are the soft Wilson lines.
Explicitly, they are defined as \cite{Bauer:2001yt,Chay:2004zn,Korchemsky:1991zp}
\begin{align}
 Y_n(x) & = \mathbf{P} \exp\left( ig_s\int^0_{-\infty}ds\, n\cdot A^a_s(x+sn)\mathbf{T}^a\right) ,\\
 Y^{\dagger}_{\bar{n}}(x) & = \mathbf{\bar{P}} \exp\left( -ig_s\int^0_{-\infty}ds\, \bar{n}\cdot A^a_s(x+s \bar{n})\mathbf{T}^a\right), \\
 Y^{\dagger}_v(x) & = \mathbf{P} \exp\left( ig_s\int_0^{\infty}ds\, v\cdot A^a_s(x+sv)\mathbf{T}^a\right)
\end{align}
with  $\mathbf{P}(\mathbf{\bar{P}})$ the (anti-)path-ordering operator acting on the color operator $\mathbf{T}^a$.
The operator $\hat{P}_k$ in Eq.(\ref{eq:soft}) is defined to extract each soft gluon's momentum.
The purpose of this paper is to calculate the soft function defined above up to NNLO.
For comparison, the threshold soft function can be obtained by changing the measurement function in Eq.(\ref{eq:soft}) to
\begin{align}
     \delta\bigg(\tau-\sum_{k}  (n \cdot \hat{P}_k+\bar{n}\cdot \hat{P}_k )/2 \bigg),
\end{align}
which has been studied before~\cite{Idilbi:2009cc,Beneke:2009rj,Czakon:2013hxa}.

\section{Renormalization} \label{sec:Renor}
Before showing the details of our calculation, it is instructive to study the structure of the soft function first,
which can guide our calculation and serve as a check of the final results.
The soft function in Eq.(\ref{eq:soft}) is defined in terms of bare parameters and
contains ultra-violate divergences when calculated in QCD perturbation theory.
These divergences can be absorbed into counterterms, leaving the renormalized soft function finite.
Following the convention, we make use of dimensional regularization, i.e., generalizing the dimension of space-time to $d=4-2\epsilon$.
After renormalization, the soft function depends on the renormalization scale $\mu$ in Eq.(\ref{eq:fact}).
This intermediate scale can appear only in logarithmic forms or in $\alpha_s(\mu)$,
the former completely determined by the corresponding counterterm.
The similar situation happens for the other components in Eq.(\ref{eq:fact}).
Since the all-order cross section does not depend on this intermediate scale,
the dependence on the scale $\mu$ will cancel against each other order by order in $\alpha_s$.
As a consequence, we can derive the logarithmic terms of the scale $\mu$ in the soft function from the knowledge of the NNLO hard and beam functions.

According to dimensional analysis, the bare soft function can be written in QCD perturbation theory as
\begin{align}\label{eq:baresoft}
     S(\tau,\mu) = \delta(\tau) + \frac{1}{\tau} \sum_{n=1}^{\infty} \left(\frac{Z_{\alpha_s} \alpha_s}{4\pi}\right)^n \left(\frac{\tau}{\mu}\right)^{-2n\epsilon} s^{(n)}~,
\end{align}
where  we use renormalized strong coupling $\alpha_s$ and its renormalization factor $Z_{\alpha_s}=1-\beta_0 \alpha_s/(4\pi \epsilon)+\mathcal{O}(\alpha_s^2)$.
The coefficients $s^{(n)}$ do not depend on $\tau$.
It is convenient to discuss the renormalization  group equation by using Laplace transformed soft function, given by
\begin{align}\label{eq:lapsoft}
    \tilde{S}(L,\mu) &= \int_{0}^{\infty} d\tau \exp\left( -\frac{\tau}{e^{\gamma_E}\mu e^{L/2}} \right) S(\tau)
          \nonumber \\
        &= 1+\sum_{n=1}^{\infty} \left(\frac{Z_{\alpha_s}\alpha_s}{4\pi} \right)^n e^{n(L-2\gamma_E)\epsilon}\Gamma(-2n\epsilon) s^{(n)}~.
\end{align}

Then the corresponding renormalized soft function $\tilde{s} $ is defined as
\begin{align}\label{eq:soft_ren}
     \tilde{s} = Z_s^{-1} \tilde{S}~,
\end{align}
where the renormalization factor $Z_s$ satisfies the differential equation
\begin{align}
    \frac{ d\ln Z_s} {d\ln\mu} = -\gamma_s
\end{align}
with $\gamma_s$  the anomalous dimension of the soft function.

If we know the result of the  anomalous dimension  $\gamma_s$,
integrating the above equation as in Refs.~\cite{Becher:2009cu,Becher:2009qa},  we  obtain a closed expression for $Z_s$
\begin{align}
     \ln Z_s &= \frac{\alpha_s}{4\pi}\left(\frac{\gamma_s^{(0)\prime}}{4\epsilon^2}+ \frac{\gamma_s^{(0)}}{2\epsilon} \right)
     + \left(\frac{\alpha_s}{4\pi}\right)^2 \left(-\frac{3 \beta_0 \gamma_s^{(0)\prime} }{16\epsilon^3} + \frac{\gamma_s^{(1)\prime} -4 \beta_0 \gamma_s^{(0)}}{16\epsilon^2}+ \frac{\gamma_s^{(1)}}{4\epsilon} \right)
     +\mathcal{O}(\alpha_s^3).
\end{align}
Here we assume the expansion series
\begin{align}
     \gamma_s = \sum_{i=0} \left( \frac{\alpha_s}{4\pi}\right)^{i+1} \gamma_s^{(i)}
     \qquad \text{and}  \qquad \gamma_s^{(i)\prime} = \frac{d \gamma_s^{(i)}}{d\ln\mu}~.
\end{align}
Now expanding Eq.~(\ref{eq:soft_ren}), we get the renormalized NLO and NNLO soft functions in Laplace space
\begin{align}
\label{eq:s_div}
    \tilde{s}^{(1)} =& \tilde{S}^{(1)}-Z_s^{(1)} ~,
    \nonumber \\
    \tilde{s}^{(2)} =& \tilde{S}^{(2)}-Z_s^{(2)} - \tilde{S}^{(1)} Z_s^{(1)}+ Z_s^{(1)2}  -\frac{\beta_0}{\epsilon} \tilde{S}^{(1)}~.
\end{align}
Since the renormalized quantities are finite, we can infer the coefficients of $\epsilon^{-i},~i=1,2,...,$ in the
bare soft function $\tilde{S}$.

In the above derivation, we have assumed the knowledge of  $\gamma_s$.
Actually, we do know it from the independence of the cross section on the renormalization scale $\mu$.
Performing Laplace transformation of Eq.(\ref{eq:fact}), we find
\begin{align} \label{eq:s_rg}
  \frac{d\ln \tilde{s}} {d\ln \mu} =\gamma_s  = -\frac{d\ln H}{d\ln \mu} - \frac{d\ln \tilde{B}_1}{d\ln \mu} - \frac{d\ln\tilde{B}_2}{d\ln \mu}~,
\end{align}
where $ \tilde{B}_i$ is the beam function after Laplace transformation.
The NLO and NNLO beam functions have already been calculated in Refs.~\cite{Stewart:2010qs,Berger:2010xi,Gaunt:2014xga,Gaunt:2014cfa}.
It is found that the beam function satisfies the same renormalization group evolution as the jet function to all orders~\cite{Stewart:2010qs}.
We write the renormalization group equation for the beam function as
\begin{align}\label{eq:beam}
     \frac{d\tilde{B}_i}{d\ln\mu} = \left(- \mathbf{T}_i\cdot \mathbf{T}_i \gcusp \Big( \ln \frac{m^2}{\mu^2} +L\Big)+ \gamma_B^i\right) \tilde{B}_i~,
\end{align}
where $\mathbf{T}_i$ is the  color generator associated with the $i$-th parton~\cite{Catani:1996jh,Catani:1996vz},
$\gcusp$ and $\gamma_B^i$ are  anomalous dimensions controlling the double and single logarithms, respectively~\cite{Gaunt:2014xga,Gaunt:2014cfa}.
The subscript or superscript $i$ can be $1$ or $2$, denoting the beam functions of the two initial-state particles, respectively.

The renormalization group equation for the hard function was studied at NNLO in Refs.~\cite{Ferroglia:2009ep,Ferroglia:2009ii}.
In our case, there is only one color basis.
It is straightforward to organise the renormalization group equation for the  hard function as
\begin{align}\label{eq:hard}
     \frac{d\ln H}{d\ln\mu} =& (\mathbf{T}_1\cdot \mathbf{T}_1 + \mathbf{T}_2\cdot \mathbf{T}_2 ) \gcusp \ln \frac{m^2}{\mu^2}
     +2\gamma^1 + 2\gamma^2 + 2\gamma^Q~.
\end{align}
The anomalous dimensions $\gamma^{1,2}$ and $\gamma^Q$,
associated with the initial- and final-state particles, can be found in Refs. \cite{Ferroglia:2009ep,Ferroglia:2009ii} and references therein.

Inserting Eqs.~(\ref{eq:s_rg}-\ref{eq:beam}) to Eq.(\ref{eq:hard}), we obtain the anomalous dimension of the soft function,
\begin{align}
     \gamma_s =& (\mathbf{T}_1\cdot \mathbf{T}_1 + \mathbf{T}_2\cdot \mathbf{T}_2)\gcusp L
     - 2 \gamma^Q
     - \gamma_B^1-\gamma_B^2 - 2\gamma^1-2\gamma^2~,
\end{align}
of which the explicit expression is available up to NNLO.

\section{NLO soft function}\label{sec:nlo}
The LO soft function is trivial and has been given explicitly in Eq.(\ref{eq:baresoft}).
In this section, we present its NLO result.

Expanding the soft Wilson lines in Eq.(\ref{eq:soft}) in series of the strong coupling, we write
the NLO soft function as
\begin{align}
   S^{(1)}(\tau) =\frac{ 2 e^{\gamma_E \epsilon}\mu^{2\epsilon} }{\pi^{1-\epsilon}}
 \int d^dq \delta(q^2) J_a^{\mu (0)\dagger}d_{\mu\nu}(q) J_a^{\nu (0)}(q)F(n,\bar{n},q)~,
\end{align}
where $e^{\gamma_E \epsilon}$ is inserted because we use $\overline{\rm MS}$ renormalization scheme.
$d_{\mu\nu}(q)$ is the polarization tensor, defined as
\begin{align}
     d_{\mu\nu}(q) = \sum_{pol.}\varepsilon_{\mu}(q)\varepsilon^*_{\nu}(q)
\end{align}
with $\varepsilon_{\mu}(q)$ the polarization vector of the soft gluon.
The factor $J_a^{\mu (0)}(q)$ is the LO one-gluon soft current, or the eikonal current,
\begin{align}
    J_a^{\mu (0)}(q) = \sum_{i=1}^{3} \mathbf{T}_i^{a} \frac{p_i^{\mu}}{p_i \cdot q}
\end{align}
with $a$ the color index.
The color conservation implies $q\cdot J_a^{(0)}(q)=0$ so that we can choose
\begin{align}\label{eq:polsum}
     d_{\mu\nu}(q) =-g_{\mu\nu}.
\end{align}
The factor $F(n,\bar{n},q)$ is a measurement function, embodying the constraint from the 0-jettiness variable.
In the center-of-mass frame, it is defined as
\begin{align}
     F(n,\bar{n},q) & =
     \delta(q^+-\tau)\Theta(q^--q^+)
      + \delta(q^--\tau)\Theta(q^+-q^-)~,
\end{align}
where we use the notations, $q^+=q\cdot n$ and  $q^-=q\cdot \bar{n}$.

After performing the phase space integration, we obtain the NLO bare soft function
\begin{align}
    s^{(1)}=\frac{4 e^{\gamma_E \epsilon}}{\epsilon \Gamma(1-\epsilon)} \bigg(\mathbf{T}_1\cdot \mathbf{T}_1 + \mathbf{T}_2\cdot \mathbf{T}_2
    +
    \mathbf{T}_3 \cdot \mathbf{T}_3 \epsilon \left( \epsilon \psi^{0}\left(1+\frac{\epsilon}{2}\right) - \epsilon \psi^{0}\left(\frac{1+\epsilon}{2}\right)-1 \right)\bigg)~,
\end{align}
where $\psi^{0}(x)$ is the digamma function which is defined as $\psi^0(x)=d\ln \Gamma(x)/dx$~.


\section{NNLO soft function} \label{sec:nnlo}

\begin{figure*}
\centering
    \centering
    \includegraphics[scale=0.5]{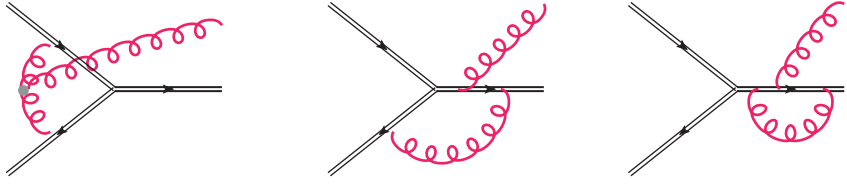}
    \caption{Selected diagrams of virtual-real contribution to the NNLO soft function.
    The double lines represent the Wilson lines and the curved lines denote the soft gluons.}
    \label{fig:virtual-real}
\end{figure*}

\begin{figure*}
\centering
    \centering
    \includegraphics[scale=0.5]{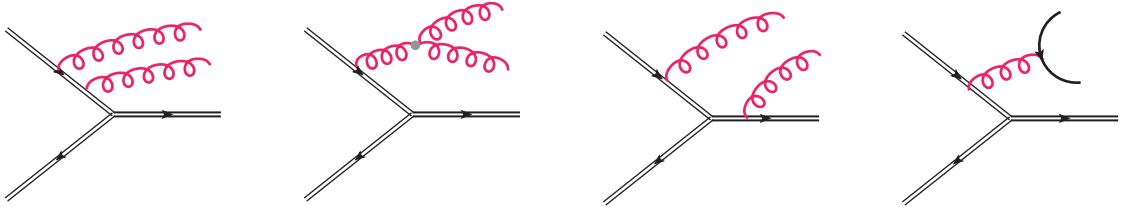}
    \caption{Selected diagrams of double-real contribution to the NNLO soft function.
    The double lines represent the Wilson lines and the curved lines denote the soft gluons. The solid lines denote the soft quark.}
    \label{fig:double-real}
\end{figure*}

Due to the interest of practical application, we consider two kinds of color structures.
One is $\mathbf{3}\otimes \mathbf{\bar{3}}\to \mathbf{8}$, denoting a massive color octet production from a triplet and an anti-triplet initial states.
The other is $\mathbf{3}\otimes \mathbf{8}\to \mathbf{3}$, representing a massive color triplet production from a triplet and an octet initial states.
We will present the result of a color octet production in the main text while leave the result of a color triplet production in the appendix.
The extension to other color structures can be obtained immediately.

The NNLO contribution consists of two parts, i.e.,
\begin{align}
     s^{(2)} = s_{\rm VR}^{(2)}+s_{\rm DR}^{(2)}.
\end{align}
The first part is the virtual-real correction, i.e., the one-loop virtual corrections to LO soft gluon current;
the second part is the double-real correction, i.e., the corrections with a double-gluon soft current or a massless quark-pair emission.
Because they have different final states, they can be calculated independently.

\subsection{Virtual-real corrections}

The virtual-real contribution to the NNLO soft function comes from the situation when two emitted soft gluons contact together to form a loop,
as  shown in Fig.~\ref{fig:virtual-real}.
In principle, one needs to calculate this kind of contribution by expanding the soft Wilson lines in Eq.(\ref{eq:soft})
and selecting the diagrams with a loop.
In practice, one can utilize the result of the soft limit of one-loop QCD amplitudes which has been studied
in Refs.~\cite{Bern:1998sc,Bern:1999ry,Catani:2000pi} and Ref.~\cite{Bierenbaum:2011gg} for massless and  massive external particles, respectively.
The unrenormalized one-loop soft current can be written as~\cite{Bierenbaum:2011gg}
\begin{align}
    J_{a}^{\mu (1)}(q)=if_{abc}\sum_{i\neq j =1}^{3}\mathbf{T}_i^b \mathbf{T}_j^c
    \bigg(\frac{p_i^{\mu}}{p_i \cdot q}-\frac{p_j^{\mu}}{p_j \cdot q}\bigg)g_{ij}(\epsilon,q,p_i,p_j)~,
\end{align}
where $f_{abc}$ are the structure constants of the group $SU_c(3)$, and the factor $g_{ij}(\epsilon,q,p_i,p_j)$ is given by
\begin{align}\label{eq:gij}
      g_{ij}(\epsilon,q,p_i,p_j) &=
       \bigg\{
      \big[ p_i\cdot q (p_j\cdot q)^2 M_1
      + \frac{1}{2} p_j \cdot q (p_i\cdot p_j p_i\cdot q -m_i^2 p_j \cdot q )M_2 + i \leftrightarrow j\big]
      \nonumber \\ &
      + [p_i \cdot p_j p_i \cdot q p_j \cdot q - m_i^2 (p_j\cdot q)^2-m_j^2 (p_i\cdot q)^2]\frac{p_i\cdot q p_j \cdot q}{p_i \cdot p_j} M_3
      \bigg\}
      \nonumber \\ &
      \times
      \frac{2p_i \cdot p_j}{m_i^2 (p_j\cdot q)^2 - 2 p_i\cdot p_j p_i \cdot q p_j\cdot q + m_j^2 (p_j\cdot q)^2}~.
\end{align}
The master integrals  $M_1$, $M_2$ and $M_3$ are collected in Appendix~\ref{app:MI_VR}.
The structure of the dependence of $J_{a}^{\mu (1)}(q)$ on  external momenta
ensures the current conservation $q\cdot J_a^{(1)}(q)=0$ and thus the choice of the gluon polarization sum in Eq.(\ref{eq:polsum})
is still valid.

The  virtual-real contribution to the soft function is given by
\begin{align}
   S^{(2)}_{\rm VR}(\tau) =\frac{4 e^{2\gamma_E \epsilon}\mu^{4\epsilon} }{\pi^{1-\epsilon}}
 \textrm{Re}\left[\int d^dq \delta(q^2)J_a^{\mu (0)\dagger}d_{\mu\nu}(q) J_a^{\nu (1)}(q)F(n,\bar{n},q)\right]~,
\end{align}
where the measurement function is the same as the NLO.
To perform the phase space integration,
we have used the package HypExp~\cite{Huber:2005yg,Huber:2007dx} to manipulate the hypergeometric functions
appearing in the intermediate steps.
The result has a closed-form expression, given by
\begin{align}
    s^{(2)}_{\rm VR} &=-\frac{8 \CA\CF}{\epsilon^3}+ \frac{8\CA^2}{\epsilon^2}
    + \frac{4 \CA}{3\epsilon}\bigg((\pi^2-6-24\ln 2)\CA
    +3\pi^2\CF \bigg)
    \nonumber \\ &
    +\frac{4 \CA}{3}\bigg(\CA(\pi^2-33\zeta_3 + 12( \ln^2 2 + 2 \ln 2) )
    +16 \zeta_3 \CF \bigg)
    \nonumber \\ &
    -\epsilon\frac{\CA }{15} \bigg[ 2 \CA \bigg(
    \pi^4 + 30 \pi^2 \big(3-4 \ln 2    + \ln^2 2\big)
    \nonumber \\ &
    -10(24+3\ln^4 2 + 72 {\rm Li}_4(1/2)-124 \zeta_3
    + 63 \zeta_3 \ln 2)
    \bigg)
    +\pi^4 \CF\bigg]
    \nonumber \\ &
    +\mathcal{O}(\epsilon^2)~.
\end{align}
We have shown explicitly the result up to $\mathcal{O}(\epsilon)$
because there is a prefactor $\tau^{-1-2n\epsilon}$ and $\Gamma(-2n\epsilon)$ in Eq.~(\ref{eq:baresoft}) and Eq.~(\ref{eq:lapsoft}), respectively.

\subsection{Double-real corrections}
We now consider the double-real contribution to the NNLO soft function.
The selected diagrams are shown in Fig.~\ref{fig:double-real}.
This part can be consider as the real correction to the LO soft current.
Instead of using SCET to write down the amplitudes for each diagram,
we make use of the results in Refs.~\cite{Catani:1999ss,Czakon:2011ve} where
the infrared behaviour of tree-level QCD amplitudes at NNLO has been analyzed.
There are two types of contributions in this part, i.e.,
the emission of two soft gluons and the emission of a soft quark-antiquark pair.
Schematically,
\begin{align}
     S^{(2)}_{\rm DR}(\tau) = S^{(2)}_{\rm gg}(\tau)+S^{(2)}_{\rm q\bar{q}}(\tau)~.
\end{align}
We will discuss them separately.

The two-gluon soft current is given by~\cite{Catani:1999ss,Czakon:2011ve}
\begin{align}\label{eq:softdrg}
     J^{\mu\nu (0)}_{a_1a_2}(q_1,q_2) &= \frac{1}{2}\bigg\{J_{a_1}^{\mu(0)},J_{a_2}^{\nu(0)} \bigg\}
     +if_{a_1a_2a_3}\sum_{i=1}^{3} \mathbf{T}_i^{a_3}
      \bigg\{ \frac{p_i^{\mu}q_1^{\nu}-p_i^\nu q_2^\mu}{q_1\cdot q_2 p_i\cdot(q_1+q_2)}
    \nonumber \\    &
      -\frac{p_i\cdot(q_1-q_2)}{2p_i \cdot (q_1+q_2)}
 \left[ \frac{p_i^\mu p_i^\nu }{p_i\cdot q_1 p_i \cdot q_2}+\frac{g^{\mu\nu}}{q_1\cdot q_2 } \right]
     \bigg\}~,
\end{align}
where we have denoted the momenta of the two gluons as $q_1$ and $q_2$, respectively.

Then we can get its contribution to the NNLO soft function,
\begin{align}
    S^{(2)}_{\rm gg}(\tau) &= \frac{2 e^{2\gamma_E\epsilon} \mu^{4\epsilon}}{\pi^{2-2\epsilon}}\int d^dq_1d^dq_2 \delta(q_1^2)\delta(q_2^2)
    \nonumber \\ & \times
 J^{\mu_1\nu_1 (0)\dagger}_{a_1a_2}(q_1,q_2) d_{\mu_1\mu_2}(q_1)d_{\nu_1\nu_2}(q_2)J^{\mu_2\nu_2 (0)}_{a_1a_2}(q_1,q_2) F(n,\bar{n},q_1,q_2)~.
\end{align}
where $F(n,\bar{n},q_1,q_2)$ is the measurement function for double-real emission,
\begin{align}\label{eq:measure}
     F(n,\bar{n},q_1,q_2) &= \delta(q_1^++q_2^+-\tau)\Theta(q^-_1-q_1^+)\Theta(q_2^--q_2^+)
     \nonumber \\ &
      +\delta(q_1^++q_2^--\tau)\Theta(q^-_1-q_1^+)\Theta(q_2^+-q_2^-)
           \nonumber \\ &
      +\delta(q_1^-+q_2^+-\tau)\Theta(q^+_1-q_1^-)\Theta(q_2^--q_2^+)
           \nonumber \\ &
      +\delta(q_1^-+q_2^--\tau)\Theta(q^+_1-q_1^-)\Theta(q_2^+-q_2^-).
\end{align}
One can see that the whole phase space is partitioned to four pieces.
Actually, since the two-gluon soft current is symmetric under the exchange of the two gluons,
one needs to calculate only two of them.

In the case of the emission of a soft quark-antiquark pair, the tree-level amplitude square is factorized as~\cite{Catani:1999ss}
\begin{align}
     |\mathcal{M}(q_1,q_2,p_1,p_2, p_3)|^2 \simeq
     (g_s^4 \mu^{4\epsilon}) \sum_{i,j=1}^{3}  \mathcal{T}_{ij}(q_1,q_2) |\mathcal{M}(p_1,p_2,p_3)|^2~,
\end{align}
where
\begin{align}\label{eq:softdrq}
     \mathcal{T}_{ij}(q_1,q_2) = - T_R  \mathbf{T}_i\cdot \mathbf{T}_j
\frac{2p_i \cdot p_j q_1 \cdot q_2+p_i\cdot(q_1-q_2) p_j \cdot(q_1-q_2)}{2(q_1\cdot q_2)^2 p_i\cdot(q_1+q_2) p_j\cdot(q_1+q_2)}
\end{align}
with the color factor  $T_R=1/2$ in the $SU_c(3)$ group.
Therefore we can obtain the corresponding contribution to the NNLO soft function as
\begin{align}
     S^{(2)}_{\rm q\bar{q}} (\tau)=\frac{4 e^{2\gamma_E\epsilon}\mu^{4\epsilon}}{\pi^{2-2\epsilon}}\int d^dq_1d^dq_2 \delta(q_1^2)\delta(q_2^2) \sum_{i,j=1}^{3}\mathcal{T}_{ij}(q_1,q_2) F(n,\bar{n},q_1,q_2)~.
\end{align}

It is convenient to perform the phase space integration in the light-cone coordinates due to the constraint from the measurement function.
In particular, we take the parametrization of the phase space as
\begin{align}
     \int d^d q = \frac{1}{2}\int d^{d-2}q_T dq^+ dq^-,
\end{align}
and express any scalar production of two momenta in terms of their components in the light-cone coordinates.
Then we insert two identities
\begin{align}
     1=\int d\tau_1 \delta(\tau_1-q_1^\pm ),\quad 1=\int d\tau_2 \delta(\tau_2-q_2^\pm )
\end{align}
to extract the contributions from the two hemispheres.
The choice of $q_i^{\pm}$, i.e., $q_i^+$ or $q_i^-$, depends on which piece in Eq.(\ref{eq:measure}) is selected.
It is easy to calculate the integration involving the $\delta$ function.
After that, we parametrize the variables $\tau_1$ and $\tau_2$ as
\begin{align}
     \tau_1 = \tau z, \quad \tau_2 = \tau (1-z).
\end{align}
Finally, the integrals we need to calculate boil down to  four-fold integrals over a unit hypercube.
In fact, there are in total over two hundreds of integrals in our calculation
because of the different kinematic structure; e.g.,
the momentum $p_i$ in Eqs. (\ref{eq:softdrg},\ref{eq:softdrq}) can be massless or massive.
Another reason is that the constraints imposed by the measurement function in Eq.(\ref{eq:measure})
hinder the application of the normal reduction method to
present a large quantity of scalar two-loop integrals in terms of a few master integrals.
For illustration, we show explicitly one example of such integrals,
\begin{align}\label{eq:example}
     I=\int_0^1 dx \int_0^1dy \int_0^1dz\int_0^1 dt
      \frac{x^{1+2 \epsilon} y^{-1+2 \epsilon} (1-z)^{-2 \epsilon } z^{-1-2 \epsilon } (1-t)^{ -\frac{1}{2}-\epsilon} t^{-\frac{1}{2}-\epsilon } }
     {\left(x^2+z-z x^2\right) \left(1-2 x y +x^2   y^2 +4 t x y \right)}.
\end{align}
For our purpose to obtain an NNLO cross section, we only need the exact results up to $\mathcal{O}(\epsilon)$
(notice the known factor $\tau^{-1-2n\epsilon}$ in the final result).
As a result, we decide to calculate these integrals numerically up to the desired precision,
as in the calculations of the dijet soft functions for $e^+ e^-$ collisions \cite{Bell:2015lsf}
and the one-jettiness soft function for proton-proton collisions \cite{Boughezal:2015eha}.

We have adopted two different methods to deal with these four-fold integrals so that they can provide a cross-check.
In the first method, we apply the Mellin-Barnes representation
and then evaluate the Mellin-Barnes integrals numerically using MB packages~\cite{Czakon:2005rk,Smirnov:2009up}.
In the other method, we make use of the SecDec~\cite{Carter:2010hi,Borowka:2015mxa} program,
which is based on the sector decomposition method \cite{Binoth:2000ps},
to calculate these integrals numerically.
For example, the integral in Eq.(\ref{eq:example}) is evaluated to be
\begin{subequations}
    \begin{align}
        I|_{\rm MB}  & = \frac{0.392699}{\epsilon^3} +\frac{1.08879}{\epsilon^2}+ \frac{4.09324}{\epsilon}
    + 15.1676  +    50.7918 \epsilon~, \\
         I|_{\rm SecDec} & = \frac{0.392695}{\epsilon^3} +\frac{1.08876}{\epsilon^2}+ \frac{4.09301}{\epsilon}
    + 15.1668  +    50.7878 \epsilon~.
    \end{align}
\end{subequations}
We find that the results from both methods agree well with each other.
The final result of the double-real contribution is
\begin{align}\label{eq:s2DR}
     s^{(2)}_{\rm DR}&=\frac{8\CA\CF-32\CF^2}{\epsilon^3}
        \nonumber \\ &
     +\frac{1}{\epsilon^2}(46.667\CA\CF
     -8\CA^2-2.667 n_f \CF)
       \nonumber \\ &
     -\frac{1}{\epsilon}(-67.226\CA\CF
     +2.667 n_f \CA -5.6423 \CA^2
     -4.444 n_f\CF
     + 263.189 \CF^2)
           \nonumber \\ &
      +(-316.07 \CA\CF -2.957 n_f \CA
     + 54.485 \CA^2 + 4.853 n_f \CF    + 641.097 \CF^2)
           \nonumber \\ &
      +\epsilon(-531.488 \CA\CF
     -2.905 n_f \CA +92.248 \CA^2
     + 10.171 n_f \CF + 874.517 \CF^2)
        \nonumber \\ &
     +\mathcal{O}(\epsilon^2) ~.
\end{align}

As mentioned below Eq.~(\ref{eq:s_div}), the coefficients of the poles in the NNLO bare soft function
can be determined from the anomalous dimensions.
Meanwhile, we have obtained the analytical result of  $s^{(2)}_{\rm VR}$ up to $\mathcal{O}(\epsilon)$ in the last subsection.
As a result, we can derive the result of $s^{(2)}_{\rm DR}$ up to $\mathcal{O}(\epsilon^0)$ without any hard calculation.
The comparison between this result and Eq.(\ref{eq:s2DR}) is presented in Tab.~\ref{tab:divergence}.
We can see that the two results agree very well, which can be considered as a nontrivial check of our calculation.

\begin{table*}
    \centering
    \begin{tabular}{c|c|c|c|c}
    \hline\hline
      & $\epsilon^{-3}$ & $\epsilon^{-2}$ & $\epsilon^{-1}$ & $\epsilon^{0}$
        \\
        \hline
        SCET prediction&-24.8889 & 96.8889 & 158.568 & 354.032  \\
         \hline
        real calculation & -24.8889 & 96.8888 & 158.577 & 353.820  \\
        \hline
        difference      & 0  & -$1\times 10^{-4}$  & 9$\times 10^{-3}$ & -0.212 \\
            \hline\hline
    \end{tabular}
    \caption{Comparison of the coefficients of $\epsilon^{-i},i=0,1,2,3,$ in $s^{(2)}_{\rm DR}$ in two different methods.
    In the first method, denoted by SCET prediction, the results are obtained from the anomalous dimensions
    and the analytical expression of  $s^{(2)}_{\rm VR}$.
    In the second method, denoted by real calculation, the results are what we calculate by Eq.(\ref{eq:s2DR}).
     Their difference is defined as the real calculation minus the SCET prediction. }
    \label{tab:divergence}
\end{table*}

\subsection{Renormalized soft function}
In the last two subsections, we have already obtained the bare soft function at NNLO. It is straightforward to perform the renormalization according to Eq.~(\ref{eq:soft_ren}). In the end, we obtain the following renormalized soft function in Laplace space for a color octet production
\begin{align}
     \tilde{s}^{(1)}&=2\CA (L-2 \ln 2) - \CF(2L^2+\pi^2),
     \nonumber \\
     \tilde{s}^{(2)}&=\CA^2 K_{AA}
     +\CA\CF K_{AF}
     + \CF^2 K_{FF}
     + \CA n_f K_{Af}
    + \CF n_f K_{Ff}
     -\frac{A_1}{4}~,
\end{align}
with
\begin{align}
K_{AA} & = \frac{17 L^2}{3} + \Big(4 \zeta_3 + \frac{98}{9}-\frac{2\pi^2}{3}-\frac{68}{3}\ln 2\Big) L + \frac{17 \pi^2}{6} ~, \nonumber \\
K_{AF} & =  -\frac{58 L^3}{9} + \Big(-\frac{134}{9}+\frac{2\pi^2}{3}+8\ln 2\Big)L^2
     + \Big(28 \zeta_3 - \frac{808}{27}-\frac{40 \pi^2}{9}\Big)L \nonumber \\
       & -\frac{568\zeta_3}{3}+\frac{4\pi^4}{9}-\frac{268 \pi^2}{27} +\frac{52 \pi^2}{3}\ln 2   ~,  \nonumber \\
K_{FF} & =   2 L^4 + 2 \pi^2 L^2 + \frac{247 \pi^4}{90} ~, \nonumber \\
K_{Af} & = -\frac{2}{3}L^2 + \Big(\frac{8}{3}\ln 2 -\frac{20}{9}\Big)L-\frac{\pi^2}{3}~, \nonumber \\
K_{Ff} & = \frac{4}{9}L^3 +  \frac{20}{9}L^2 + \Big(\frac{112}{27}+\frac{4\pi^2}{9} \Big)L + 8\zeta_3
    + \frac{40}{27}\pi^2   ~,
\end{align}
and $A_1$ is the coefficient of $\epsilon$ in $s_{\rm DR}^{(2)}+s_{\rm VR}^{(2)}$.
Notice that the scale dependence is encoded in the variable $L$.
The above equation is the main result in this work.
It is ready  to extend to other color structures.

\section{Application to fixed-order calculations}\label{sec:app}
Given that the $N$-jettiness subtraction has proven to be successful in performing NNLO differential calculations,
we anticipate that the soft function we just obtained can be used in fixed-order calculations.
In this section, we take the  massive color-octet vector boson production at the LHC as an example,
but the method is general.

In fixed-order calculations, the cross section is divided to two parts
\begin{align}\label{eq:fixedxs}
     d\sigma = d\sigma(\tau<\taucut)+d\sigma(\tau\geq \taucut ),
\end{align}
where $\taucut$ is an intermediate cutoff parameter.
When $\taucut$ is small compared to the typical scale of the process, such as the mass of the massive color octet,
the first part can be approximated very well by Eq.(\ref{eq:sigma}).
The beam functions have been already known up to NNLO~\cite{Stewart:2010qs,Gaunt:2014xga,Gaunt:2014cfa}
and the soft function is obtained up to NNLO in this work.
The hard function is available  only up to NLO \cite{Chivukula:2011ng}.
Therefore, we can calculate this part using these functions in effective field theory.
The second part, due to the constraint $\tau>\taucut$, corresponds to the process of a color octet associated with a jet production.
We only need its cross section at (N)LO for the calculation of $d\sigma$ in Eq.(\ref{eq:fixedxs}) at (N)NLO.
As a result, it can be computed with various existing packages, e.g., MadGraph5\_aMC@NLO~\cite{Alwall:2014hca}.

\begin{figure}
\centering
    \centering
    \includegraphics[scale=0.40]{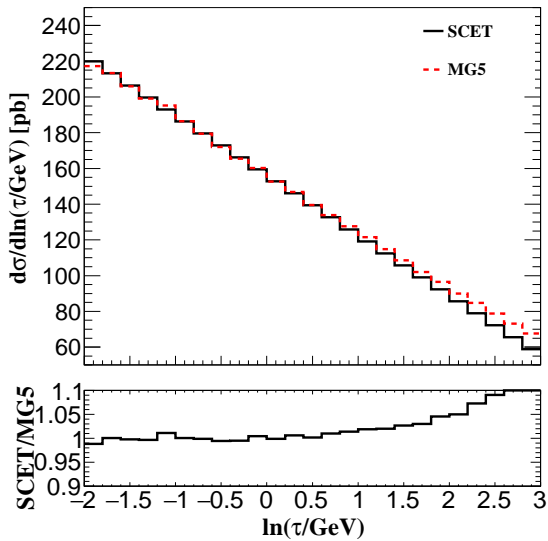}
    \caption{The NLO $\tau$ distributions from SCET and MadGraph5\_aMC@NLO.}
    \label{fig:taucut}
\end{figure}

The detailed discussion of the higher-order effects in a colored particle production using the $N$-jettiness subtraction deserves another paper,
which we leave to future work.
Instead, we now show a spectrum of the $\tau$ distribution when $\tau$ is small in Fig.~\ref{fig:taucut}.
We consider a color octet vector boson production at the 13 TeV LHC.
It has a mass of 1 TeV and couplings $r_L = r_R = 1$, which are  defined in Ref.~\cite{Chivukula:2011ng}.
The renormalization and factorization scales are both chosen at 1~TeV,
and the CT14NLO \cite{Dulat:2015mca} parton distribution function  is used.
We provide the results calculated by Eq.~(\ref{eq:fact}) in SCET and
the automatic package MadGraph5\_aMC@NLO with the model generated by  FeynRules~\cite{Alloul:2013bka}.
From Fig.~\ref{fig:taucut}, we observe that they are coincident with each other if $\tau < 1$ GeV,
which means that the prediction of the effective field theory reproduces the full QCD fixed-order calculation in the limit $\tau \to 0$,
as we just discussed.

Notice that though the spectrum shows that the differential cross section becomes very large when $\tau$ decreases,
it does not mean the prediction for an observable is infinite.
Actually, there is another large contribution, which has an opposite sign,
from the term with $\delta(\tau)$ in each order of $\alpha_s$, as shown in Eq.(\ref{eq:baresoft})
after  the expansion
\begin{align}
     \frac{1}{\tau}\left(\frac{\tau}{\mu}\right)^{\eta}=\frac{1}{\eta}\delta(\tau)+\left( \frac{1}{\tau} \right)_{+}+
     \eta \left( \frac{\ln \tau / \mu }{\tau} \right)_{+}+\cdots .
\end{align}
The plus distribution is defined as
\begin{align}
     \int_0^{\mu} d\tau \left( \frac{1}{\tau} \right)_{+} f(\tau) = \int_0^{\mu} d \tau \frac{f(\tau )-f(\tau=0)}{\tau}.
\end{align}
If an observable is infrared safe, i.e., not sensitive to the soft and collinear radiations,
then one needs to integrate over the region near $\tau=0$, e.g., from $0$ to $\Delta(>0)$,
and thus the final result is finite.

\section{Conclusions}
\label{sec:conc}
Precise theoretical predictions are important for both exact estimate of the backgrounds and the search for new physics.
To obtain reliable predictions, QCD higher-order effects need to be taken into account.
$N$-jettiness subtraction is one of the efficient methods to compute the fully differential cross section at NNLO.
In this work, we calculate one of the indispensable ingredients in $N$-jettiness subtraction method,
i.e., the soft function, for one massive colored particle production at hadron colliders.
The NLO soft function is obtained to all orders in $\epsilon$ while
the NNLO soft function is only calculated up to $\mathcal{O}(\epsilon^0)$ for our purpose.
We have used Mellin-Barnes integral representation and sector decomposition method to derive the above results.
The structure of the soft functions is in coincidence with the expectation from analysing the scale independence of the cross section.
We present the final results for the color octet in the main text and color triplet in the appendix, respectively.
After that, we briefly discuss its application to the fixed-order calculations.
For illustration, using this soft function along with other functions already known before,
we reproduce the behavior of the NLO cross section for a  massive color-octet vector boson production when
the 0-jettiness variable $\tau$ is small.
If the proper hard function is available, we can calculate the differential cross sections at higher orders.
Our method can also be used to calculate the soft function for single top or top quark pair productions.

\section*{Acknowledgements}
We would like to thank Martin Beneke for carefully reading the manuscript.
The work of HTL was supported by the ARC Centre of Excellence for Particle Physics at the Tera-scale.
The work of JW was supported by the  Cluster of Excellence
{\it Precision Physics, Fundamental Interactions and Structure of Matter} (PRISMA-EXC 1098) when
he was in Johannes Gutenberg University,
and by the BMBF project No. 05H15W0CAA at TU Munich.
HTL would like to acknowledge the Mainz Institute for Theoretical Physics for its hospitality at the beginning of this work.

\appendix
\section{Soft functions for a color triplet particle production}
\label{app:nnlo_soft}
In this appendix,
we collect the renormalized soft functions for a direct top production from quark-gluon initial states via a flavor-changing neutral current.
Since all the master integrals are the same as in the case of a color octet production,
we omit the intermediate details and show the final results directly.

The NLO soft function is
\begin{align}
     \tilde{s}^{(1)} = -(\CA+\CF)L^2 + 2 \CF L
     - \frac{1}{2}\left[ (\CA+\CF)\pi^2 + 8 \ln 2 ~\CF \right]~.
\end{align}
The NNLO soft function is
\begin{align}
     \tilde{s}^{(2)} & = \CA^2 J_{AA}
   + \CF^2 J_{FF}
 + \CA \CF J_{AF}
   + \CA n_f J_{Af}
   + \CF n_f J_{Ff}
   -\frac{\tilde{A}_1}{4}
\end{align}
with
\begin{align}
J_{AA} & = \frac{L^4}{2}-\frac{11 L^3}{9}+\left(\frac{5 \pi ^2}{6}-\frac{67}{9}\right) L^2
     + \left(14 \zeta_3 -\frac{404}{27}-\frac{11 \pi ^2}{9}\right)L
          \nonumber \\ &  -22 \zeta_3
   +\frac{109 \pi^4}{120}-\frac{134 \pi ^2}{27} ~, \nonumber  \\
J_{FF} & =
   \frac{L^4}{2}-2 L^3
   + \left(2+\frac{\pi ^2}{2}+4\ln 2\right) L^2- \left(\pi
   ^2+8\ln 2\right) L
   \nonumber \\ &
   -\frac{218 \zeta_3}{3}+\frac{247 \pi ^4}{360}+\pi ^2+\frac{26}{3}
   \pi ^2 \ln 2     ~,  \nonumber  \\
J_{AF} & =
   L^4-\frac{29 L^3}{9} +  \left(-\frac{34}{9}+\frac{4 \pi ^2}{3}+4\ln 2\right) L^2
   \nonumber \\&
   +
   \left(18 \zeta_3-\frac{110}{27}-\frac{26 \pi ^2}{9}-\frac{44 \ln2}{3}\right) L
   \nonumber \\ &
   -\frac{284 \zeta_3}{3}+\frac{287 \pi ^4}{180}-\frac{169 \pi
   ^2}{54}+\frac{26}{3} \pi ^2 \ln 2 ~, \nonumber  \\
J_{Af} & =
\frac{2 L^3}{9}+\frac{10 L^2}{9}+ \left( \frac{2 \pi ^2}{9}+\frac{56 }{27} \right) L +4 \zeta_3
   +\frac{20
   \pi ^2}{27}  ~,  \nonumber  \\
J_{Ff} & =
\frac{2 L^3}{9}+\frac{4 L^2}{9}+ \left( \frac{2 \pi ^2 }{9}-\frac{4 }{27}
   +\frac{8}{3}  \ln 2 \right) L +4 \zeta_3+\frac{11 \pi ^2}{27} ~,
\end{align}
and
\begin{align}
   \tilde{A_1} = 116.458 \CA^2 +75.764 \CF^2+149.400 \CA  \CF
   + 5.358 \CA n_f +  6.292 \CF n_f~.
\end{align}

\section{Master integrals in virtual-real corrections}
\label{app:MI_VR}

The master integrals in Eq.~(\ref{eq:gij}) are defined as~\cite{Bierenbaum:2011gg}
\begin{align}
      M_1 =& \Phi \int \frac{d^dk}{i (2\pi)^d}\frac{1}{k^2(k+q)^2 (-p_j\cdot k)}~,
      \nonumber \\
      M_2 =& \Phi \int \frac{d^dk}{i (2\pi)^d}\frac{1}{k^2(p_i\cdot k + p_i \cdot q)^2 (-p_j\cdot k)}~,
      \nonumber \\
      M_3 =& \Phi \int \frac{d^dk}{i (2\pi)^d}\frac{1}{k^2(k+q)^2(p_i\cdot k + p_i \cdot q)^2 (-p_j\cdot k)}~
\end{align}
with $\Phi=8\pi^2 (4\pi)^{-\epsilon} e^{\gamma_E \epsilon }$~.

For $p_i^2=0$ and $p_j^2=0$, they are given by
\begin{align}
    M_1 =& 0, \qquad M_2 =0,
    \nonumber
    \\
    M_3 =&\frac{2^{-\epsilon}\pi e^{ \gamma_E \epsilon}
    (p_i\cdot p_j)^{\epsilon}}{(p_i\cdot q)^{1+\epsilon} (p_j\cdot q)^{1+\epsilon}}
(-1-i\delta)^{-\epsilon} {\rm cot}(\pi \epsilon) \Gamma(-\epsilon)\Gamma(1+2\epsilon),
\end{align}
where $\delta\to 0^+$ is introduced to indicate the way of continuation.

For $p_i^2=m_i^2$ and $p_j^2=0$, they are given by
\begin{align}
     M_1 &= 0~,
     \nonumber \\
     M_2 &= 2^{-2\epsilon}e^{ \gamma_E\epsilon} (-1-i \delta)^{-2\epsilon} \Gamma(-\epsilon)\Gamma(2\epsilon) m_i^{2\epsilon} \frac{(p_i\cdot q )^{-2\epsilon}}{ p_i\cdot p_j} ~,
          \nonumber \\
     M_3 &= -\frac{2^{-1-\epsilon}e^{\gamma_E\epsilon} (p_i \cdot p_j )^{\epsilon}}{(p_i\cdot q)^{1+\epsilon} (p_j \cdot q)^{1+\epsilon}} \Gamma(-\epsilon) \Gamma(2\epsilon)\bigg\{
     2(-1-i \delta)^{-2\epsilon}~_2F_1(-\epsilon, -\epsilon,1-\epsilon,x)
     \nonumber \\ &
     + \frac{\Gamma(1-2\epsilon)\Gamma(1+\epsilon)}{\Gamma(1-\epsilon)}
     ~_2F_1(-\epsilon,1+\epsilon,1-\epsilon,x)
     \bigg\}
\end{align}
with
\begin{align}
     x=1-\frac{m_i^2 p_j\cdot q}{2 p_i\cdot p_j p_i\cdot q}~.
\end{align}

For $p_j^2=m_j^2$ and $p_i^2=0$, they are given by
\begin{align}
     M_1 &= -2^{-1-2\epsilon}e^{\gamma_E\epsilon} (-1-i \delta)^{-2\epsilon} \Gamma(-\epsilon)\Gamma(2\epsilon) \frac{m_j^{2\epsilon}}{(p_j\cdot q)^{1+2\epsilon}},
     \nonumber \\
     M_2 &= \pi e^{\gamma_E\epsilon}\csc(2\pi \epsilon) \Gamma(\epsilon)m_j^{-2\epsilon}\frac{(p_i\cdot q)^{-2\epsilon} }{(p_i\cdot p_j)^{1-2\epsilon} }~,
     \nonumber \\
     M_3 &= M_3(p_i^2=m_i^2, p_j^2=0)|_{i\leftrightarrow j}~.
\end{align}

\bibliography{main-JHEP}
\bibliographystyle{JHEP}

\end{document}